\documentclass[twocolumn]{aastex701}

\accepted{September 26, 2025}

\shorttitle{LATIS Data Release}
\shortauthors{Newman et al.}

\begin{document}

\title{LATIS Data Release: $\sim4200$ Spectra of $z \sim 2-3$ Galaxies, Redshifts, and Intergalactic Medium Tomography Maps}

\correspondingauthor{Andrew B. Newman}
\email{anewman@carnegiescience.edu}

\author[0000-0001-7769-8660]{Andrew B. Newman}
\affiliation{Observatories of the Carnegie Institution for Science, 813 Santa Barbara St., Pasadena, CA 91101, USA}
\email{anewman@carnegiescience.edu}

\author[0000-0002-8459-5413]{Gwen C. Rudie}
\affiliation{Observatories of the Carnegie Institution for Science, 813 Santa Barbara St., Pasadena, CA 91101, USA}
\email{gwen@carnegiescience.edu}

\author[0000-0003-4218-3944]{Guillermo A. Blanc}
\affiliation{Observatories of the Carnegie Institution for Science, 813 Santa Barbara St., Pasadena, CA 91101, USA}
\affiliation{Departamento de Astronomía, Universidad de Chile, Camino del Observatorio 1515, Las Condes, Santiago, Chile}
\email{gblanc@carnegiescience.edu}

\author[0000-0003-4727-4327]{Daniel D. Kelson}
\affiliation{Observatories of the Carnegie Institution for Science, 813 Santa Barbara St., Pasadena, CA 91101, USA}
\email{kelson@carnegiescience.edu}

\author[0000-0003-3691-937X]{Nima Chartab}
\affiliation{Observatories of the Carnegie Institution for Science, 813 Santa Barbara St., Pasadena, CA 91101, USA}
\affiliation{Caltech/IPAC, 1200 E. California Blvd., Pasadena, CA 91125, USA}
\email{nchartab@ipac.caltech.edu}

\author{Enrico Congiu}
\affiliation{European Southern Observatory (ESO), Alonso de Córdova 3107, Casilla 19, Santiago 19001, Chile}
\email{econgiu@eso.org}

\author{Victoria P\'{e}rez}
\affiliation{Departamento de Astronomía, Universidad de Chile, Camino del Observatorio 1515, Las Condes, Santiago, Chile}
\email{victoriapaz.perezgonzalez@gmail.com}

\author[0000-0001-7066-1240]{Mahdi Qezlou}
\affiliation{The University of Texas at Austin, 2515 Speedway Boulevard, Stop C1400, Austin, TX 78712, USA}
\email{qezlou@austin.utexas.edu}

\author[0000-0001-5803-5490]{Simeon Bird}
\affiliation{Department of Physics and Astronomy, University of California Riverside, 900 University Ave., Riverside, CA 92521, USA}
\email{sbird@ucr.edu}

\author[0000-0002-1428-7036]{Brian C. Lemaux}
\affiliation{Gemini Observatory, NSF NOIRLab, 670 N. A’ohoku Place, Hilo,
HI 96720, USA}
\affiliation{Department of Physics and Astronomy, University of California, Davis, One Shields Ave., Davis, CA 95616, USA}
\email{brian.lemaux@noirlab.edu}

\author[0000-0002-9336-7551]{Olga Cucciati}
\affiliation{INAF - Osservatorio di Astrofisica e Scienza dello Spazio di Bologna, via Gobetti 93/3, 40129 Bologna, Italy}
\email{olga.cucciati@inaf.it}

\begin{abstract}
We present the data release of the Ly$\alpha$ Tomography IMACS Survey (LATIS), one of the largest optical spectroscopic surveys of faint high-redshift galaxies. The survey provides 7408 optical spectra of candidate $z\sim2$--3 galaxies and QSOs in the Canada--France--Hawaii Telescope Legacy Survey D1, D2 (COSMOS), and D4 fields. The $R \sim 1000$ spectra were obtained using the Inamori Magellan Areal Camera and Spectrograph (IMACS) at the Magellan Baade telescope, with typical integrations of 12~hr. From these spectra, we measured 5575 high-confidence spectroscopic redshifts, of which 4176 are at $z > 1.7$, thereby substantially increasing the number of public spectroscopic redshifts at $z \approx 2$--3 in COSMOS and the other survey fields. The data release includes Ly$\alpha$ transmission fluctuations measured in $4.7 \times 10^5$ pixels, which were used to create 3D maps of the intergalactic medium (IGM) transmission spanning 1.65~deg${}^2$ and $z = 2.2$--2.8 at a resolution of 4~$h^{-1}$~cMpc. These are the largest such maps to date and provide a novel tracer of large-scale structure in legacy fields. We also provide ancillary data, including mock surveys. The LATIS data will enable a variety of community studies of galaxy evolution, environments, and the IGM around cosmic noon.
\end{abstract}


\section{Introduction}
\label{sec:intro}

The Ly$\alpha$ Tomography IMACS Survey (LATIS) was designed to enable $\sim$Mpc resolution mapping of the intergalactic medium (IGM) at $z \sim 2.5$ by observing the Ly$\alpha$ forest in the spectra of faint galaxies \citep{Lee14a}. The motivation, survey design, target selection, observations, data reduction,  and the main analysis methods for LATIS were described by \citet{Newman20}. The LATIS data have already been used to detect protoclusters and study galaxy evolution within them \citep{Newman22,Qezlou22,Newman25a,Newman25b}, to measure the relationship between the matter density and IGM transmission \citep{Newman25b}, and to constrain the stellar mass--halo mass relation \citep{Newman24}, the stellar mass--stellar metallicity relation \citep{Chartab24}, and the star formation--density relation \citep{Chartab25} at $z \sim 2.5$.

LATIS is one of the largest optical (rest-frame far-ultraviolet) spectroscopic surveys of faint ($r < 24.8$) high-redshift galaxies, and the data can support many uses by the community. In this paper we describe and document the data release, which includes 7408 spectra, 5575 high-confidence redshift measurements, Ly$\alpha$ forest transmission fluctuation in $4.7 \times 10^5$ pixels, and IGM tomography maps covering a volume of $4 \times 10^6$~$h^{-3}$~cMpc${}^3$. We also provide ancillary data that are useful for analyzing these products.

As in previous LATIS work, we use the \citet{Planck15} cosmological parameters.

\begin{figure*}
    \centering
    \includegraphics[width=\linewidth]{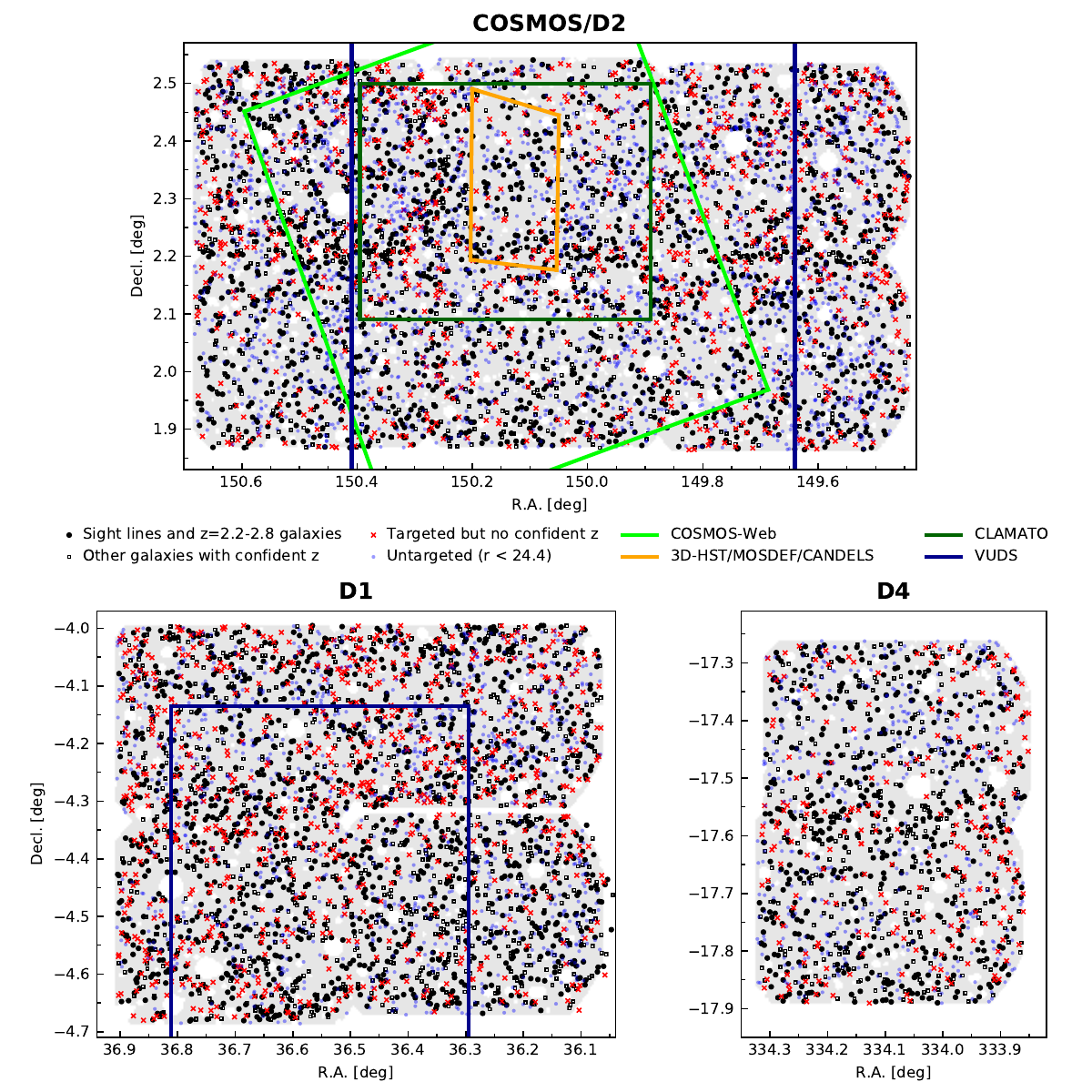}
    \caption{Maps of the LATIS survey fields. Filled circles show targets that were either confirmed to lie in the targeted redshift range $z=2.2$--2.8, or were used as Ly$\alpha$ forest sight lines to map absorption in that range. Open squares show other galaxies for which high-confidence redshifts were obtained. Red crosses show galaxies for which we were unable to measure a high-confidence redshift. Blue circles show unobserved galaxies in the parent sample to a limit of $r < 24.4$, which were given top priority; many fainter galaxies ($r = 24.4$--24.8) are present in the parent sample and were targeted at lower priority, but these are not shown for clarity. The gray background shows the survey area, comprised of the union of the IMACS footprints with regions around bright stars omitted. (A few galaxies in the southwest of the D1 field lie outside this nominal survey area, because the mask center was shifted after initial observations to capture better guide stars.) The approximate footprints of the COSMOS-Web NIRCam \citep{Casey23}, 3D-HST \citep{Brammer12}, CLAMATO DR2 \citep{Horowitz22}, and VUDS \citep{LeFevre15} surveys are shown; the MOSDEF \citep{Kriek15} and CANDELS \citep{Grogin11,Koekemoer11} footprints are similar to 3D-HST. Overlap with other surveys is illustrated in Fig.~1 of \citet{Newman20}.}
    \label{fig:fields}
\end{figure*}

\section{Observations}
\label{sec:obs}

All LATIS observations were conducted with the Inamori Magellan Areal Camera and Spectrograph (IMACS; \citealt{Dressler11}) at the Magellan Baade telescope. We used the f/2 camera, which, given our mask design constraints, provided an effective field of view of 0.15~deg${}^2$. To improve blue sensitivity and multiplexing, we used a custom grism and bandpass filter \citep{Newman20}. The bandpass filter isolates the wavelength range 387--586~nm with $>$95\% transmission; the spectra were ultimately trimmed to the range 389--583~nm. The grism provides a dispersion of 1.7--2.0~\AA~per (binned) spectral pixel, which is resampled to 1.8~\AA~per pixel in our reduced spectra. For the average observed image size, the spectral resolution is well approximated by $\sigma_{\rm inst} = [227 - 98 \times \lambda / (500~\textrm{nm})]$~km~s${}^{-1}$, where $\lambda$ is the observed wavelength. This corresponds to a resolving power that ranges from $R = 840$ to 1140 over the filter bandpass. Slits were $1\farcs2$ wide with a minimum length of $6''$. 

LATIS observations were obtained in the Canada--France--Hawaii Telescope Legacy Survey (CFHTLS) fields D1, D2, and D4. The CFHTLS D2 field is within COSMOS. Figure~\ref{fig:fields} outlines the surveyed regions. The areas covered by slit masks are 0.559~deg${}^2$, 0.824~deg${}^2$, and 0.271~deg${}^2$ in D1, D2, and D4, respectively, for a total of 1.654~deg${}^2$. Excluding the regions around stars that were omitted from target selection reduces these areas to 0.533~deg${}^2$, 0.767~deg${}^2$, and 0.255~deg${}^2$ in D1, D2, and D4, respectively, for a total of 1.555~deg${}^2$.

As discussed by \citet{Newman20}, the survey area comprises 12 IMACS ``footprints,'' and we typically observed two distinct target sets per footprint. Furthermore, since exposures of a given target set were split over multiple observing runs, we were usually able to identify targets outside the desired redshift range and swap them for new candidates.

The photometric parent sample, the construction of slit masks, and the prioritization of targets were described in detail by \citet{Newman20}; here we briefly summarize. The Lyman-break galaxy (LBG) sample comprises 98\% of targets. In every survey field, candidate LBGs were selected using $ugr$ color criteria developed to isolate galaxies at $z \approx 2.2$--3.2. In COSMOS, we additionally selected candidate LBGs with photometric redshifts $z_{\rm phot} = 2.2$--3.2.\footnote{The selection methods have substantial overlap, as expected: 50\% of LBG candidates in COSMOS were selected by both methods, 33\% were only $ugr$-selected, and 17\% were only $z_{\rm phot}$-selected. Among the targets that we ultimately confirmed to lie at $z > 2.2$, 78\% were selected by both methods, 7\% were only $ugr$-selected, and 15\% were only $z_{\rm phot}$-selected.} We also observed known quasars (QSOs) and candidates that were selected based on their $ugr$ colors and photometric variability. Literature spectroscopic redshifts $z_{\rm spec}$ were used to include targets in the desired redshift range and to exclude candidates known to lie outside that range; however, 86\% of targets were uninformed by a literature $z_{\rm spec}$. The highest priority was given to quasars, LBGs with known redshifts, and candidate LBGs in the prime magnitude range $r_{\rm min} < r < 24.4$, where $r_{\rm min} = 23.5$ for the $ugr$-selected candidates and $r_{\rm min} = 23.0$ for the $z_{\rm phot}$-selected candidates. Secondary priority was given to brighter or fainter candidates in the range $r = 23.0$--24.8.

In addition, we observed a series of bright-target masks in suboptimal conditions of seeing or transparency. The primary targets on these masks span magnitudes $r = 22.0$--23.5, but some ``filler'' targets at fainter magnitudes were also included. We did not observe the entire survey area with bright-target masks---all of the D1 field was observed, but only one-third of COSMOS and none of D4---and observations of these masks did not reach a uniform sensitivity. We therefore recommend excluding galaxies that were observed only on the bright-target masks for applications that require spatial uniformity (i.e., galaxy density maps).

Observations were obtained over a total of 59 operable nights between 2017 December and 2022 August. (Four nights in 2017 September were used for pilot observations.) We cut 71 slit masks. In many cases, separate masks were cut for observations east and west of the meridian to reduce the effects of differential atmospheric refraction.\footnote{The ADC used with IMACS removes the wavelength-dependent atmospheric refraction, but not the achromatic variation in scale along the parallactic direction.} The masks thus comprise 53 distinct sets of targets. 

\begin{figure*}
    \centering
    \includegraphics[width=6in]{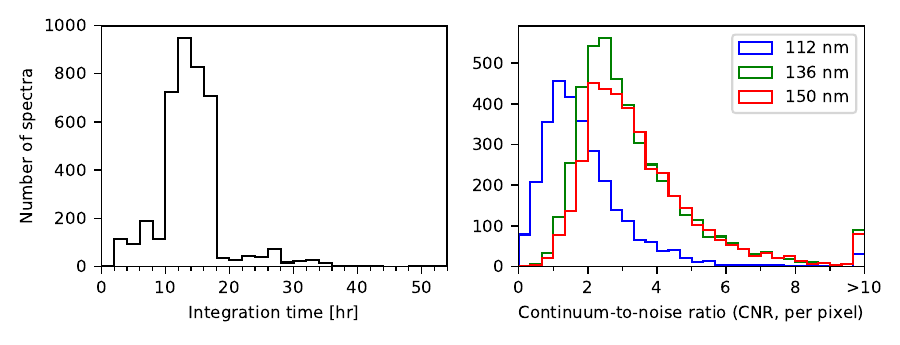}
    \caption{\emph{Left panel:} The distribution of total integration times for targets with high-confidence redshifts $z > 2$. \emph{Right panel:} The distribution of the median CNR per spectral pixel for the same targets, evaluated in 40~\AA~intervals centered on the rest-frame wavelengths indicated in the caption, which were chosen to be free of strong absorption or emission lines.}
    \label{fig:exptimedist}
\end{figure*}

We cut slits for a total of 7972 targets and obtained a reduced spectrum for 7408 targets, of which 6568 were observed on regular masks and 840 were observed only on bright-target masks. Nearly all of the slits lacking a spectrum are explained by gaps in the detector mosaic. Figure~\ref{fig:exptimedist} (left panel) shows the distribution of exposure times. The median integration time is 12.1~hr, but it ranges from 1.5--52.8~hr depending on the number of slit masks on which a target was included. The right panel of Figure~\ref{fig:exptimedist} shows the distribution of the continuum-to-noise ratio (CNR) for targets found to be at redshifts $z > 2$. (The CNR matches the common signal-to-noise ratio for the bins around rest-frame wavelengths of 136~nm and 150~nm, but for the 112~nm bin in the Ly$\alpha$ forest, we use the unabsorbed continuum.)

\section{Data Reduction and Validation}

The data reduction procedure was described by \citet{Newman20}. Here we describe several improvements that are reflected in the current data release, and have been used in \citet{Newman24} and all subsequent LATIS papers.

\subsection{Broadband Flux Calibration}

A flux calibration curve was derived from observations of a small number of white dwarf standards \citep{Moehler14} through a wide slit obtained during each observing run (see Figure~6 of \citealt{Newman20}). We built up a global calibration curve by averaging those obtained over the survey, and for each run, we modeled the calibration curve as the product of the global curve and a linear function of wavelength. We therefore kept the higher-order features of the calibration curve consistent for the entire data set, while also allowing for low-order time variation due to various factors, e.g., mirror reflectivity. 

The standard star spectra were always dispersed over the same two detectors of the eight-detector mosaic. We added an additional step to correct for differences in the wavelength-dependent sensitivity of the detectors. (This is not accomplished by our flat-fielding step, which corrects only for pixel-to-pixel response differences and illumination variations along a slit.) We did so by analyzing the relative shapes of the twilight flat spectra on all detectors. As a test, we then performed the same calculation using the average spectrum (in the observed frame) of the high-$z$ galaxies; this comparison suggests that broadband relative flux calibration is good to $\simeq 5$\%.

\emph{We made no attempt to correct for approximately gray effects on the flux normalization due to slit losses, non-photometric conditions, vignetting, or Galactic extinction.} The LATIS survey fields have low reddening $E(B-V) = 0.016$--0.025~mag \citep{SF11}, which corresponds to a differential extinction of only 0.03--0.04~mag between the blue and red ends of the LATIS spectra. Our spectral modeling (Section~\ref{sec:spectral_modeling}) uses a power law to modulate the continuum shape, which easily absorbs this very small reddening. 

\emph{The absolute fluxes in the LATIS spectra are therefore not meaningful.} Comparison to $g$-band photometry (corrected for Galactic extinction) indicates that the LATIS spectra have lower flux densities by a factor $2.1 \pm 0.5$ (median and standard deviation).

\subsection{Narrowband Flux Calibration}

\begin{figure}
    \centering
    \includegraphics[width=3in]{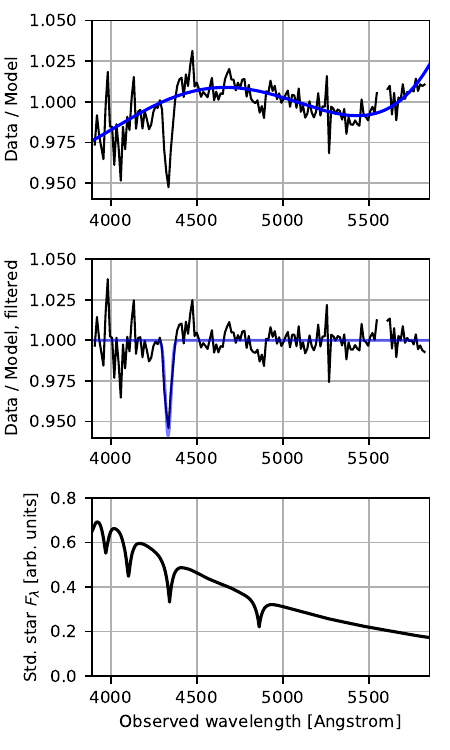}
    \caption{\emph{Top panel:} Average ratio of observed and model spectra for $z=1.8$--3.0 LBGs, in the observed frame. The Ly$\alpha$ forest and strong interstellar absorption lines are excluded from each spectrum. A polynomial fit is shown in blue. \emph{Middle panel:} Same as the top panel, but with the polynomial fit divided out. A Gaussian model of the systematic residual centered at 4332~\AA~is overlaid; this was used to correct the flux calibration curve derived from white dwarf standards. \emph{Bottom panel:} White dwarf standard spectrum demonstrating that the feature arose due to mismatch at H$\gamma$.}
    \label{fig:fluxcalsmall}
\end{figure}

To examine the accuracy of the relative flux calibration on smaller wavelength scales, we examined  residuals between LBG spectra and the model fits described in Section~\ref{sec:spectral_modeling}. The top panel of Figure~\ref{fig:fluxcalsmall} shows the average ratio of observed to model spectra in the observed frame. The Ly$\alpha$ forest was excluded so that deviations cannot be caused by large-scale structure. Furthermore, narrowband deviations are unlikely to be produced by errors in the models, which average out when combining spectra in the observed frame. However, model errors may cause broadband deviations, so we divided out the low-order features to produce the middle panel of Figure~\ref{fig:fluxcalsmall}.

Any narrowband residuals imply an error in the global flux calibration curve. We found one clear residual (see middle panel of Figure~\ref{fig:fluxcalsmall}), which is well fit by a Gaussian centered on 4332~\AA~with a width of $\sigma = 16$~\AA~and an amplitude of 5.9\%. This indicates that the flux calibration procedure did not remove the intrinsic white dwarf spectrum accurately around the H$\gamma$ absorption line. We modified the global flux calibration curve accordingly, by dividing it by this Gaussian model. (The residual corresponds to $z_{\rm Ly\alpha} = 2.563$, which resulted in enhanced absorption at this redshift in the version of the maps analyzed by \citealt{Newman22}, as discussed by \citealt{Newman25b}.)

\subsection{Atmospheric Dispersion Corrector Failures}
\label{sec:adc} 

Six slit masks were observed, at least in part, without a functioning atmospheric dispersion corrector (ADC), which resulted in curved spectral traces and a loss of light at blue wavelengths. For these masks, we modeled the curved traces and extracted spectra along a curved aperture. We estimated the light loss by comparing the average galaxy spectrum (in the observed frame) on a given mask to the average galaxy spectrum over all masks within the same survey field that were observed with a functional ADC. We found a selective light loss of $\lesssim 30$\% at blue wavelengths relative to the red. We fit polynomials to estimate a relative flux calibration correction appropriate to each affected slit mask. 

\subsection{Sky Subtraction}

We corrected a bias in the sky subtraction that particularly affected blue wavelengths.\footnote{The problem was traced to a biased estimator used to identify outliers in the background spectrum.} Sky subtraction is challenging due to the short slits (typically 6~arcsec) used on LATIS slit masks to maximize the sight-line density for IGM tomography. We evaluated the magnitude of any remaining systematic sky subtraction errors in two ways. First, we constructed a composite spectrum of LATIS LBGs and compared it to the \citet{Reddy16} composite spectrum. We focused on the depth of the higher-order Lyman series lines, beneath which Reddy et al.~detected residual light and inferred a non-unity covering fraction of neutral gas. The average redshift of LATIS galaxies that contribute to the stack around Ly$\beta$ is $z \approx 3$, close to the Reddy et al.~mean redshift. We found a reasonable agreement and estimated that any errors are $\lesssim 4\%$ of the flux density in the Ly$\alpha$ forest. Second, \citet{Newman24} showed that the agreement between galaxy--Ly$\alpha$ and galaxy--galaxy clustering implies that multiplicative errors in the Ly$\alpha$ transmission fluctuations $\delta_F$, which would be induced by commensurate additive errors in the spectra, are $\lesssim 4$\%.

Any global biases in sky subtraction should therefore be $\lesssim 4$\% of the median flux density in the Ly$\alpha$ forest, or $\simeq$0.012~$\mu$Jy (in the flux units of the spectra, which are not corrected for slit losses). \emph{We do not recommend LATIS spectra for applications that are very sensitive to a pedestal of this magnitude, such as Lyman continuum studies.} Furthermore, Ly$\alpha$ emission is often spatially extended and can leak into the background spectrum that we subtract, which can bias both the Ly$\alpha$ flux and equivalent width. Our data reduction did not attempt to mitigate this effect, so \emph{this limitation should be kept in mind for any studies that involve Ly$\alpha$ emission.}

\section{Spectral Classification, Modeling, and Redshift Measurements}

The data reduction resulted in 7408 spectra, calibrated to vacuum wavelengths and $f_{\nu}$ flux densities. Here we review the classification and modeling of the spectra and the resulting data products.

\subsection{Visual Classification and Masking}

One of us (A.B.N.) visually reviewed the 1D and 2D spectra for every target. We interactively assigned a spectral type, corresponding to a template spectrum. We used SDSS templates that include QSOs, low-$z$ galaxies, and stars\footnote{\url{https://classic.sdss.org/dr2/algorithms/spectemplates/index.php}}, as well as the \citet{Shapley03} LBG composite spectrum. An initial redshift was assigned interactively and then refined by cross-correlation with the template. For the LBGs and QSOs, the redshift was ultimately determined by spectral modeling as discussed below; for the low-$z$ galaxies (nearly always $z < 0.5$), this cross-correlation provided the final redshift estimate. During the inspection, we assigned various flags, as discussed below, and comments.

An automatic procedure initially flagged corrupted wavelength regions by detecting anomalies in the background spectra. We refined the mask interactively. Reasons for masking include contamination by the zeroth-order image of another slit, bad sky subtraction, second-order light from an alignment star, detector gaps, and reduction errors. The bright 5577~\AA~sky line was always masked.

\subsection{Spectral Modeling}
\label{sec:spectral_modeling}

Spectra of LBGs and QSOs were modeled as detailed by \citet{Newman20}. We have retained the same procedure for the QSOs modeling but slightly adjusted the LBG modeling, as we summarize here.

We iteratively refined the LBG redshifts and a set of template spectra bootstrapped from LATIS itself. Initial redshifts were estimated using the \citet{Shapley03} template. We then built a set of template spectra (see below). Each spectrum was modeled as the redshifted product of a non-negative linear combination of these templates, modulated by a power law and the mean transmitted flux $\langle F(z) \rangle$ in the Ly$\alpha$ forest. The newly derived redshifts were then used to construct  new templates, and the fitting was repeated; two iterations were sufficient to converge.

The LBG templates were constructed as the average of LATIS spectra in bins of Ly$\alpha$ equivalent width, which is a proxy for the spectral morphology, after dividing out a power-law continuum and $\langle F(z) \rangle$ for each galaxy (see \citealt{Newman20} for details). The systemic redshift of the templates was anchored to the \ion{C}{3} $\lambda1175.71$ stellar photospheric line. During the survey, we recognized and corrected a subtle bias in the template construction. Due to the restricted wavelength range over which we observe, the mean redshift of the galaxies that contribute to an average spectrum varies significantly with wavelength. The average spectral slope can vary with redshift, due to evolution and/or our color selection, which can in turn affect the template continuum shape. We found evidence for such an effect by detecting systematic residuals from the model fits when averaged in the rest frame. We fit these residuals with smooth functions and modified the template continuum shapes accordingly.

For purposes of analyzing the Ly$\alpha$ forest, we adjusted our model of the unabsorbed continuum for each galaxy using a low-order multiplicative polynomial. This method, known as mean flux regularization (MFR; \citealt{Lee12}), forces the Ly$\alpha$ transmitted flux $F$ to match the mean flux $\langle F \rangle(z)$ when averaged over large scales $\Delta z \simeq 0.2$ (see Section 7.1 of \citealt{Newman20}).

\subsection{Redshift Measurements, Quality Flags, and Reliability}
\label{sec:z}

The LATIS redshift measurements, quality flags, and other information are tabulated in a FITS table distributed with this data release. The contents of this catalog are described here and in the data release documentation.

The main targets of interest for LATIS are the LBGs and QSOs, whose redshifts were determined by the modeling procedure described above and in \citet{Newman20}. For some target, which have the {\tt manualz} flag set to True in the catalog, the redshift was not determined in this way. Such classes include (1) all low-redshift galaxies, whose redshifts were instead determined by cross-correlation with the SDSS templates; (2) stars, whose redshifts are all set to 0; (3) seven broad absorption line (BAL) QSOs, which contain ``BAL'' in the {\tt comments} field and have very uncertain redshifts; (4) type 2 active galactic nuclei (AGN) distinguished by the presence of high-ionization emission lines like \ion{He}{2} and \ion{C}{4}.

For the 50 AGN, which contain ``AGN'' in the {\tt comments} field, most redshifts were determined by interactively measuring the wavelengths of emission lines. We corrected the resulting redshifts by adding a velocity $\Delta v$ as determined by \citet{Hainline11}. We used \ion{He}{2} when observed ($\Delta v = 0$), and generally Ly$\alpha$ ($\Delta v = -197$~km~s${}^{-1}$) otherwise; if Ly$\alpha$ had multiple peaks, we used \ion{C}{4} or \ion{Si}{4} ($\Delta v = -273$~km~s${}^{-1}$). In some cases, the AGN features were weak enough that we could model the spectrum as an LBG; these have {\tt manualz} set to False.

\begin{table*}[]
    \centering
    \begin{tabular}{c|c|c|c}
         {\tt zqual} & Description & Frac. of spectra & Confidence level \\\hline
         0 & No redshift measured & 14\% & $\ldots$ \\
         1 & Only a single emission line was detected & 0.7\% & $\ldots$ \\
         2 & Low confidence & 10\% & $\gtrsim 50\%$ \\
         3 & High confidence & 20\% & $\gtrsim 94$\% \\
         4 & Very high confidence & 56\% & $\gtrsim 99$\%       
    \end{tabular}
    \caption{Spectroscopic redshift quality flags}
    \label{tab:zqual}
\end{table*}

\begin{figure*}
    \centering
    \includegraphics[width=6in]{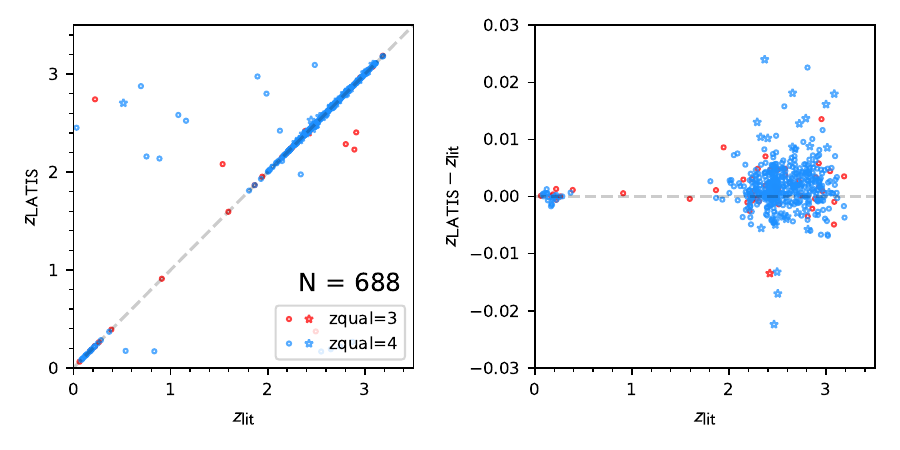}
    \caption{Comparison of LATIS redshifts $z_{\rm LATIS}$ to literature redshifts $z_{\rm lit}$ in the COSMOS compilation of \citet{Khostovan25} and VUDS. See the text for discussion of the quality flags used. Circles and stars include LATIS LBGs and QSOs, respectively.}
    \label{fig:zcomp}
\end{figure*}

We assigned a quality flag {\tt zqual} to each spectrum based on our subjective confidence its reliability (Table~\ref{tab:zqual}). We estimate the error rate in each {\tt zqual} class by cross-matching to independent measurements from several sources. We included spectroscopic redshifts from a COSMOS compilation \citep{Khostovan25}, considering only those with a quality flag 3 or 4 (or 13 or 14, to include broad-line spectra) and therefore a stated confidence level $\geq 95$\%. To these we added redshifts from the VIMOS Ultra-Deep Survey (VUDS; \citealt{LeFevre15}) using the catalog of \citet{Lemaux22}, again taking sources with high quality flags XX3 and XX4. Figure~\ref{fig:zcomp} compares the LATIS and literature redshifts for 688 sources in common.

Among {\tt zqual} $ = 4$ spectra of LBGs, 2.9\% (16 of 550 matches) were initially identified as outliers with redshift differences $> 1000$~km~s${}^{-1}$. Reviewing the LATIS spectra, we found that the LATIS redshift is unambiguous in 11 cases, while in 5 cases the LATIS measurement is more ambiguous and likely a lower {\tt zqual} should have been assigned. In no case was the literature $z$ clearly indicated in the LATIS spectrum. We estimate an error rate of $\lesssim 5/550 = 0.9$\%. This is an upper limit because some of the literature redshifts are also expected to be incorrect.

Among {\tt zqual} $ = 3$ spectra, 16\% (15 of 95 matches) were initially identified as outliers. Upon reviewing the LATIS spectra, we concluded that the LATIS redshift was correct in 9 cases, the literature redshift was correct in 2, and 4 were ambiguous. We estimate that the error rate is $\lesssim 6 / 95 = 6$\%. 

Only 18 of our {\tt zqual} $ = 2$ spectra have a high-confidence redshifts from the other catalogs, and 10 are outliers. Most of the LATIS redshifts are ambiguous, suggesting the error rate might be as high as $9 / 18 = 50\%$. This is supported by anomalies in the average spectrum of the {\tt zqual} $ = 2$ sources. Only a very small number of galaxies were classified with {\tt zqual} $ = 1$, so we cannot quantify a confidence level. 

\emph{In this data release, we include only redshifts with {\tt zqual} = 3 or 4 and set the others to NaN.} No LATIS analysis has ever used these low-confidence redshifts ({\tt zqual} $ = 1$ or 2), which comprise only 11\% of spectra, and we consider them to be unsuitable for science applications. Spectra of all sources are included in the data release. Based on the above estimates, the catastrophic  error rate is $\lesssim 2$\% for the distributed redshifts ({\tt zqual} $\geq 3$).

Among the LATIS QSOs, we find 43 matches to the other redshift catalogs, of which 2 differ by $> 5000$~km~s${}^{-1}$ from the LATIS measurement. Both are BAL QSOs. We consider BAL redshifts to be very approximate and do not use them for any of our analyses. 

Based on the above comparisons, we modified the redshifts of three sources for the final released catalog.

We note that the identification of LBG redshifts in LATIS did not rely strongly on Ly$\alpha$ emission, and we do not expect there to be a significant selection effect favoring targets with higher rest-frame Ly$\alpha$ equivalent width (EW$_0$). The targets show a wide range of EW$_0$, and the average spectrum has a mix of broad Ly$\alpha$ absorption and modest emission (see Figures 8 and 15 of \citealt{Newman20}). Absorption dominates in 61\% of targets (${\rm EW}_0 < 0$) and only 10\% of targets would be classified as Ly$\alpha$ emitters (${\rm EW}_0 > 20$~\AA).\footnote{For our purposes, which are only indicative, we integrated from $-1000$ to +2000~km~s${}^{-1}$ of Ly$\alpha$ at the systemic redshift when measuring EW$_0$, which is roughly $\pm 1500$~km~s${}^{-1}$ of the average peak position. We defined the continuum by linearly interpolating the average flux densities within the two continuum wavelength intervals used by \citet{Kornei10}.}

\subsection{Random and Systematic Redshift Errors}

Rest-UV redshifts generally show offsets and scatter from the systemic values due to gas flows and resonant scattering. Systematic errors in the LATIS LBG redshifts were evaluated by \citet{Newman24}, who found that adding $\Delta v = 39 \pm 16$~km~s${}^{-1}$ to the redshifts was necessary to symmetrize the Ly$\alpha$ absorption observed in transverse sightlines. \emph{We have added this correction to the model-based LBG redshifts in the catalog distributed with this paper.} 

We estimated random errors by comparing to near-infrared (NIR) and radio observations in the \citet{Khostovan25} compilation, which should better trace the systemic redshift with small errors. Excluding QSOs and AGN, we found a normalized median absolute deviation (NMAD) of 89~km~s${}^{-1}$, which we estimate to be the random uncertainty in the LATIS LBG redshifts. The median offset of $\Delta v = -14 \pm 18$~km~s${}^{-1}$, after applying the correction described in the last paragraph, supports the accuracy of that correction.

We note that the LATIS redshifts are slightly higher than the average literature redshift (Figure~\ref{fig:zcomp}), most of which are derived from rest-UV spectra. These tests show that the LATIS redshifts are closer to systemic, which is important when correlating galaxies with IGM absorption.

QSO redshifts have much higher uncertainties. Only optical redshifts are available for comparison in the \citet{Khostovan25} catalog. We find an NMAD velocity difference of $\approx 800$~kms${}^{-1}$. If we consider the literature and LATIS redshifts to have similar and uncorrelated random errors, this suggests uncertainties of about 600~km~s${}^{-1}$ in each. 

\subsection{Other Flags and Comments}

The catalog contains several Boolean flags, which are set to true under the following conditions:

\begin{itemize}
    \item {\tt manualz}: The redshift was not determined by QSO or LBG spectral fitting, as discussed above.
    \item {\tt isqso}: The target was modeled using the broad-line QSO templates.
    \item {\tt isstar}: The target was classified as a star.
    \item {\tt badredux}: The spectrum was affected by serious reduction issues that compromise the whole spectrum and so cannot be solved by masking (4\% of spectra). This often happens due to serendipitous sources in the slit affecting the sky subtraction, which may be noted in the comments. These spectra are not recommended for detailed analysis or inclusion in stacks, but a redshift might still be obtained. 
    \item {\tt tomo}: The spectrum was included in the IGM tomography analysis. Among sources in the relevant redshift range, 8\% were excluded; reasons include {\tt badredux}, {\tt manualz}, a poor model fit, or multiple galaxies identified in the same spectrum.
    \item {\tt brionly}: The target was observed only on bright-target masks.
    \item {\tt adc\_corr}: Corrections for a malfunctioning ADC (Section~\ref{sec:adc}) were applied to some of the exposures of this target.
\end{itemize}

The comments field includes freeform remarks and may also contain one or more standardized phrases:

\begin{itemize}
    \item \emph{``AGN'' or ``BAL'' (57 sources)}. Described above.
    \item \emph{``Blended spectra'' (199 sources)}. Inspection of the light profile along the slit, as well as images, suggests that a second source might contribute significant light. In most cases, these spectra look normal. However, users may wish to exclude these spectra for some applications.
    \item \emph{``Two galaxies'' (52 sources).} Two galaxy spectra are clearly blended as indicated by, e.g., multiple Ly$\alpha$ lines or a second full suite of interstellar absorption lines. The spectrally dominant source is assigned {\tt z}, while the second redshift may be manually estimated and recorded in the {\tt z2} field.
    \item \emph{``Strong Lya'' (40 sources)}. The Ly$\alpha$ emission line is significantly stronger than the highest-EW LBG template. These often require a significant MFR adjustment to fit the Ly$\alpha$ forest reasonably.
\end{itemize}

\begin{table*}[]
    \centering
    \begin{tabular}{lccc}
         Sample & Number on main & Number on bright-target & Sum \\
         & masks & (backup) masks & \\\hline
         Total spectra & 6568 & 840 & 7408 \\
         Confident $z_{\rm spec}$ (zqual = 3 or 4) & 5156 & 419 & 5575 \\
         Galaxies at $z > 1.7$ & 4019 & 56 & 4075 \\
         QSOs at $z > 1.7$ & 96 & 5 & 101 \\
         Galaxies at $z < 0.5$ & 1029 & 357 & 1386 \\
         Stars & 356 & 181 & 537
    \end{tabular}
    \caption{Breakdown of target numbers}
    \label{tab:targnums}
\end{table*}

For spectra in which we recognized possible foreground gas absorption via multiple \ion{H}{1} or metal absorption lines, we recorded an estimate of the redshift in the {\tt z2} field. We do not guarantee any level of completeness or purity for these identifications; we used them only to mask metal lines at the same redshift in the Ly$\alpha$ forest for our IGM tomography analysis.

\section{Spectroscopic Sample: Yields, Targeting Rates, and Success Rates}

\begin{figure}
    \centering
    \includegraphics[width=3.5in]{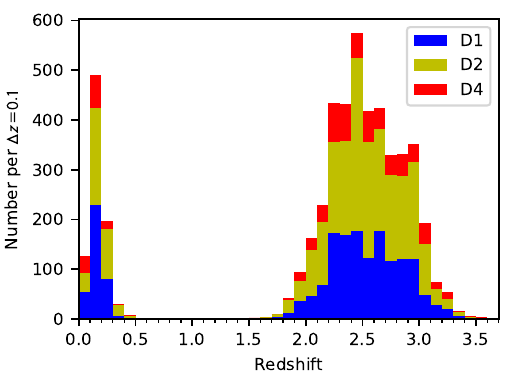}
    \caption{The distribution of confident ({\tt zqual} = 3 or 4) galaxy redshifts measured in each LATIS survey field. }
    \label{fig:zdist}
\end{figure}

\begin{figure*}
    \centering
    \includegraphics[height=6in]{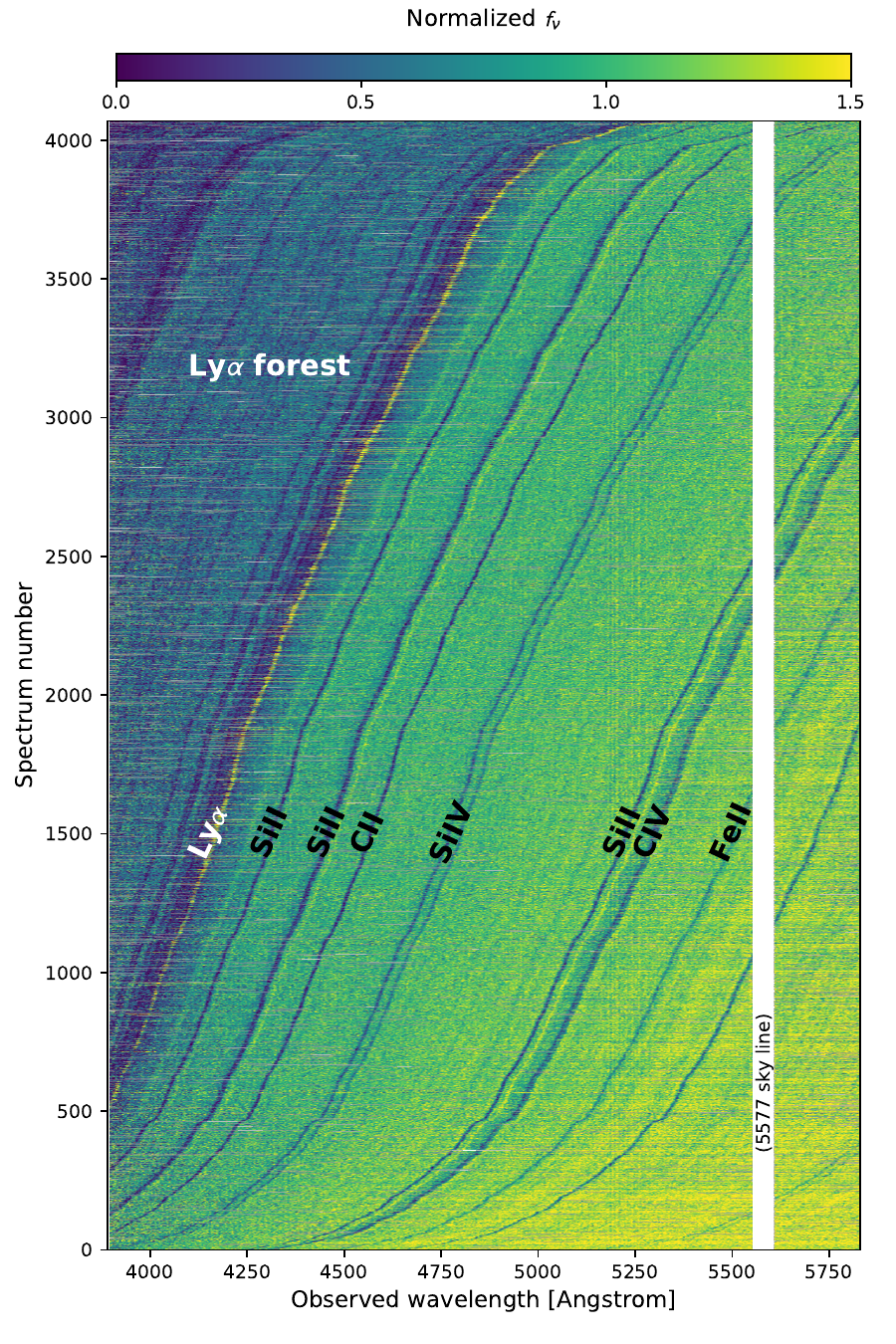}
    \caption{Montage of LATIS spectra of LBGs and QSOs at $z > 1.7$. Spectra are ordered in redshift from bottom to top and normalized by their median $f_{\nu}$ over $\lambda_{\rm rest} = 1300$--1500~\AA.}
    \label{fig:montage}
\end{figure*}

In total we obtained 5575 confident redshifts (defined as {\tt zqual} = 3 or 4), whose distribution is shown in Figure~\ref{fig:zdist}. Most are in the high-redshift peak, with 4176 LBGs and QSOs at $z > 1.7$. Figure~\ref{fig:montage} graphically illustrates the spectra of these sources. Individual examples can be found in Figure~8 of \citet{Newman20}. 

The main ``failures'' are $z < 0.5$ galaxies (1386) and stars (537), which occur disproportionately in the bright-target masks. The breakdown of target classifications is listed in Table~\ref{tab:targnums}. The bimodal redshift distribution arises from two sources: (1) the main failure mode of the color and $z_{\rm phot}$ selection criteria is confusion between the Lyman and Balmer breaks, and (2) the observed bandpass is not rich in strong spectral features for redshifts between the two peaks, making identification difficult.

\subsection{Targeting and Spectroscopic Success Rates}

\begin{figure}
    \centering
    \includegraphics[width=\linewidth]{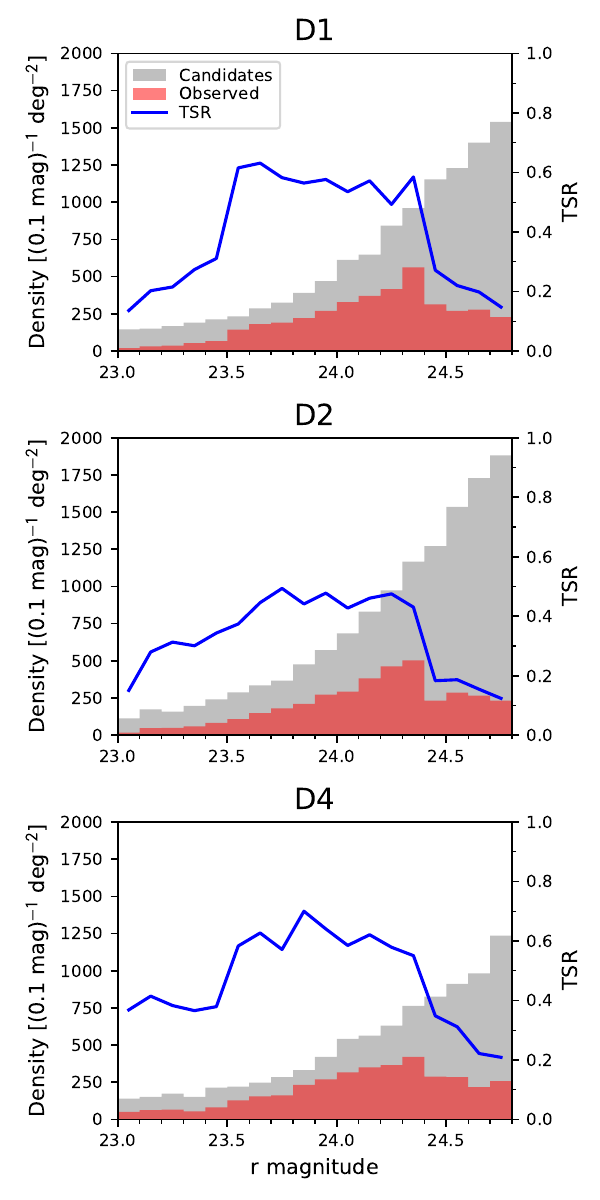}
    \caption{Targeting statistics per field. The gray histogram shows the surface density of LBG candidates in the parent catalog, as a function of $r$ magnitude. The subset targeted in LATIS is shown by the red histogram. The ratio of the two histograms is the TSR (blue curve, right axis). Updated from Figure~11 of \citet{Newman20}.}
    \label{fig:tsr}
\end{figure}

The target sampling rate (TSR) is the fraction of objects in the parent catalog for which spectra were obtained by LATIS. Figure~\ref{fig:tsr} shows that the TSR is relatively flat in the prime magnitude range $r = 23.5$--24.4 and declines for brighter and fainter sources, as expected from the prioritization scheme discussed in Section~\ref{sec:obs}. In the prime magnitude range, the TSR is approximately 60\% in the D1 and D4 fields and 45\% in the D2 field. The lower TSR in D2 (COSMOS) is a consequence of the larger parent catalog, produced by the addition of $z_{\rm phot}$-selected targets.

\begin{figure}
    \centering
    \includegraphics[width=3in]{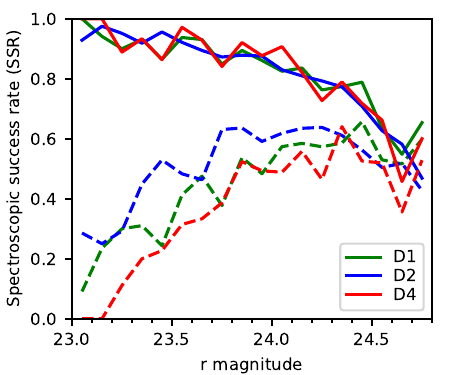}
    \caption{The spectroscopic success rate per field. Solid curves show the fraction of targeted sources for which a high-confidence redshift was determined. Dashed curves show the fraction of targeted sources with a high-confidence redshift in the desired range, $z_{\rm spec} = 2.2$--3.2.}
    \label{fig:ssr}
\end{figure}

The spectroscopic success rate (SSR) is the fraction of targeted objects for which a confident redshift was determined. The SSR increases for brighter objects (Figure~\ref{fig:ssr}), as expected, and reaches nearly unity at the bright limit $r = 23$. However, if we consider successes to be only confident redshifts in the range to which the selection criteria were tuned, $z = 2.2$--3.2, then the success rate falls for brighter objects, due to the increased contamination by interlopers. 

The TSR and SSR also depend on the angular position within the survey field due to a variety of factors, including variations in observing conditions, variations in image quality within the IMACS FOV, and a tendency for slits to be placed at the top and bottom of a mask more often than the center to maximize the number of slits. These factors were modeled by \citet{Newman24} in their Appendix B. We distribute maps of the relative TSR and SSR with this data release. The maps are intended to be applied only to the main sample, i.e., excluding sources with the {\tt brionly} flag set. We computed the TSR maps for the subset of targets in the magnitude range $r = r_{\rm min}$--24.8 (see Section~\ref{sec:obs}) plus the QSOs. Although the absolute value of the TSR varies strongly with magnitude (Figure~\ref{fig:tsr}), for many applications (e.g., galaxy clustering and overdensities) only the relative angular variation is required.

Finally, \citet{Newman22} investigated the sampling of close pairs. They found a dearth, relative to the parent catalog, of pairs of LATIS targets separated by $< 12''$, which is an inevitable consequence of the $6''$ slit length and two-pass observing strategy. This limitation is important to keep in mind for clustering studies.

\section{IGM Tomography}

We measured the Ly$\alpha$ forest transmission fluctuations
\begin{equation}
    \delta_F = \frac{F}{\langle F(z) \rangle} - 1,
\end{equation}
where $F = S / C$ is the transmitted flux, $S$ is the observed spectrum, and $C$ is the unabsorbed continuum model. The procedure was detailed by \citet{Newman20}. $C$ was determined from our spectral modeling including MFR (Section~\ref{sec:spectral_modeling}). For each $\delta_F$ measurement, we computed the random uncertainty $\sigma_{\rm noise}$, an estimate of the correlated uncertainty $\sigma_{\rm cont}$ arising from misplacement of the continuum (see \citealt{Newman20}, Section 7.4), and a total uncertainty $\sigma_{\rm tot} = (\sigma_{\rm noise}^2 + \sigma_{\rm cont}^2)^{1/2}$. 

Pixels in the Ly$\alpha$ forest were masked for several reasons: (1) The spectrum itself may be masked due to data reduction issues. (2) Several of the stronger absorption lines in the intrinsic spectrum of the background galaxy were masked, as listed by \citet{Newman20}. (3) In cases where foreground metal absorption at a redshift {\tt z2} was identified redward of Ly$\alpha$, we masked $\pm 300$~km~s${}^{-1}$ intervals around the wavelengths of a suite of metal lines at {\tt z2} that fell within the Ly$\alpha$ forest. (4) We identified and masked lines with high equivalent widths, with the aim of excluding damped Ly$\alpha$ and other high column density absorption lines that are produced by dense gas rather than diffuse IGM. The algorithm was described in Section 3.1 of \citet{Newman24}. (5) We omitted 1\% of pixels with a very low CNR, i.e., ${\rm CNR} = C / \sigma < 0.5$, where $\sigma$ is the random uncertainty in $S$, and a handful of pixels with extreme values $|\delta_F| > 10$.

Among the LATIS spectra, we selected 3052 sight lines (1070, 1574, and 408 in fields D1, D2, and D4, respectively) for which the Ly$\alpha$ forest (1040--1187~\AA) probes foreground absorption at $z = 2.2$--2.8, which implies a redshift $z = 2.277$--3.442. These have the {\tt tomo} flag set to True. In a few spectra, the entire Ly$\alpha$ forest is masked, resulting in 3012 sight lines containing a total of 469,008 pixels in which $\delta_F$ was measured.

\begin{figure}
    \centering
    \includegraphics[width=3.5in]{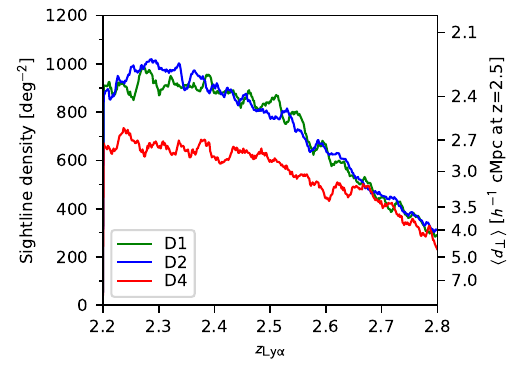}
    \caption{The areal density of sight lines sampling the Ly$\alpha$ forest at a redshift $z_{\rm Ly\alpha}$. The right axis shows the mean transverse separation $\langle d_{\perp} \rangle$ evaluated in $h^{-1}$ cMpc at $z = 2.5$. (The conversion from the left to right axis varies with redshift only by $\lesssim 6$\%.)}
    \label{fig:sld}
\end{figure}

The effective sight-line density $A$ is shown in Figure~\ref{fig:sld}. This quantity measures the areal density of sight lines that probe the Ly$\alpha$ forest at a given $z_{\rm Ly\alpha}$. The total density of all sight lines is much higher (1845 deg${}^{-2}$ averaged over all survey fields), because the Ly$\alpha$ forest in a given sight line spans only a portion of the total redshift range. The effective sight-line density can also be expressed in terms of a mean transverse separation $\langle d_{\perp} \rangle = (70.6~h^{-1}~{\rm cMpc})  \times (A / {\rm deg}^{-2})^{-1/2}$, where the conversion is evaluated at $z = 2.5$. The design goal of LATIS was $\langle d_{\perp} \rangle \lesssim 3~h^{-1}$~cMpc \citep{Newman20}, which is met for $z \approx 2.2$--2.65, while $\langle d_{\perp} \rangle$ increases to $\approx4$~$h^{-1}$~cMpc at $z=2.8$. Although we built IGM maps over $z = 2.2$--2.8, we note that the LATIS spectra also probe Ly$\alpha$ absorption at $z > 2.8$ with a relatively low sight-line density; however, they cannot probe significantly below $z=2.2$ due to the filter cutoff and the sensitivity of IMACS.

\begin{figure}
    \centering
    \includegraphics[width=3.5in]{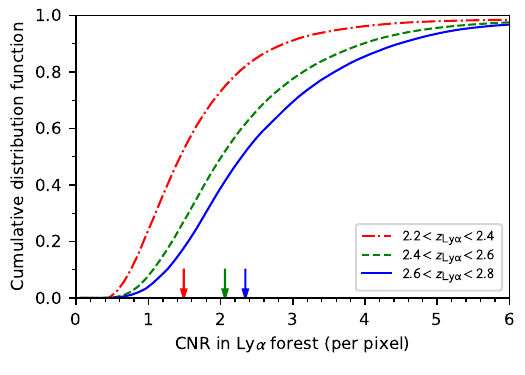}
    \caption{Cumulative distributions of the continuum-to-noise ratio (per pixel) in the Ly$\alpha$ forest within three intervals of $z_{\rm Ly\alpha}$. Vertical arrows indicate median values.}
    \label{fig:cnr}
\end{figure}

The distribution of CNR in the Ly$\alpha$ forest is shown in Figure~\ref{fig:cnr}. The LATIS design goal was ${\rm CNR} = 2$ in 1.8~\AA~pixels. The median achieved values were 1.5, 2.1, and 2.3 in the redshift ranges $z_{\rm Ly\alpha} = 2.2$--2.4, 2.4--2.6, and 2.6--2.8, respectively.

\subsection{IGM Maps}

To produce the IGM tomography maps, we first converted the ($\alpha$, $\delta$, $z$) coordinates, where $\alpha$ and $\delta$ represent the R.A.~and decl.~in degrees, respectively, of each spectral pixel into the ($X$, $Y$, $Z$) coordinates of the map. In this coordinate system, $X$ increases to the west, $Y$ to the north, and $Z$ with redshift:
\begin{eqnarray}
X &=& -(\alpha - \alpha_0) \cos(\delta_0)  D_C(z) \pi/180 + \Delta X / 2 \nonumber\\
Y &=& (\delta - \delta_0) D_C(z) \pi/180 + \Delta Y / 2\nonumber\\
Z &=& D_C(z) - D_C(2.2)
\label{eqn:coords}
\end{eqnarray}
Here $D_C(z)$ is the comoving distance in units of $h^{-1}$~cMpc. The central celestial coordinates are $(\alpha_0, \delta_0) = (36.4749, -4.3384)$ for D1, $(150.0606, 2.2026)$ for D2, and $(334.0852, -17.5797)$ for D4. The transverse map dimensions in $h^{-1}$~cMpc are $(\Delta X, \Delta Y) = (65, 53)$ for D1, $(93, 51)$ for D2, and $(33, 47)$ for D4. Note that $D_C(z)$ depends on $z$, so the sight lines flare outward in the map volume. The total volume enclosed by the sight lines is $3.96 \times 10^6$~h${}^{-3}$~cMpc${}^3$. This is smaller than the rectangular volume due to the sight-line flaring, i.e., the full transverse area is not sampled by sight lines at the low-$z$ end of the maps.

Each spectral pixel in the Ly$\alpha$ forest can then be described by coordinates ($X$, $Y$, $Z$), the measured $\delta_F$, and its uncertainty, for which we use $\sigma_{\rm tot}$. In each survey field, we provided these data to the {\tt dachshund} code \citep{Stark15}, which produces a Wiener-filtered map with (1~$h^{-1}$~cMpc)${}^3$ voxels as its output. The parameters governing the Wiener filter, specifically the signal covariance, were described in Section 8.2 of \citet{Newman20}. Finally we smoothed the Wiener filter output by an isotropic 3D Gaussian with $\sigma = 4$~$h^{-1}$~cMpc.

We note that the maps are in redshift space. Furthermore, {\tt dachshund} places the \emph{half-integer} coordinates at the centers of voxels. Thus, to compute the ($\alpha$, $\delta$, $z$) coordinates of the center of the lower-left voxel, one should use $(X, Y, Z) = (0.5, 0.5, 0.5)$ in Equation~\ref{eqn:coords}.

The data release provides smoothed maps of $\delta_F / \sigma_{\rm map}$, where $\sigma_{\rm map}$ is the standard deviation of the map, excluding voxels within 4~$h^{-1}$~cMpc of a boundary to mitigate edge effects. For the D1, D2, and D4 fields, $\sigma_{\rm map} = 0.0496$, 0.0476, and 0.0458, respectively. These maps were first provided by \citet{Newman25a}; we include them in this data release along with the other products. We also provide maps of $\delta_F$ that are not normalized by $\sigma_{\rm map}$. Section~\ref{sec:mocks} provides critical information needed to interpret these values. Noise in the maps was estimated using mock surveys; see \citet{Newman25a}, in particular the last paragraph of their Section 2.

\subsection{Mock Surveys and Calibration of Transmission Fluctuations}
\label{sec:mocks}

The data release provides our mock surveys performed within the MultiDark Planck 2 (MDPL2) simulation snapshot at $z=2.535$ \citep{Klypin16}. These mocks have been described in previous LATIS papers \citep{Newman20,Newman22,Newman25a}. In brief, the Ly$\alpha$ optical depth was computed using the fluctuating Gunn--Peterson approximation (FPGA; \citealt{Gunn65,Croft98,Weinberg99}). For each survey field, we selected 100 subvolumes of the (1~$h^{-1}$~Gpc)${}^3$ box. Each subvolume was sampled by skewers matching the relative positions of the actual LATIS sight lines. Mock spectra were generated matching the spectral resolution and noise characteristics of the LATIS data, including continuum errors. The processing of the spectra and construction of the maps followed the procedures applied to the LATIS observations, including smoothing of the Wiener filter output. This resulted in a total of 300 mock maps, whose values we denote by $\delta_F^{\rm rec}$. In addition, for each subvolume we computed a noiseless estimate $\delta_F^{\rm true}$ by directly convolving a dense grid of sight lines (spaced by 0.25~$h^{-1}$~cMpc) by a $\sigma = 4$~$h^{-1}$~cMpc kernel, without any Wiener filtering. 

\begin{figure*}
    \centering
    \includegraphics[width=3.5in]{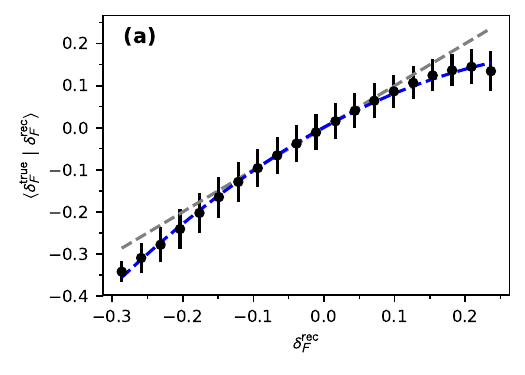}
    \includegraphics[width=3.5in]{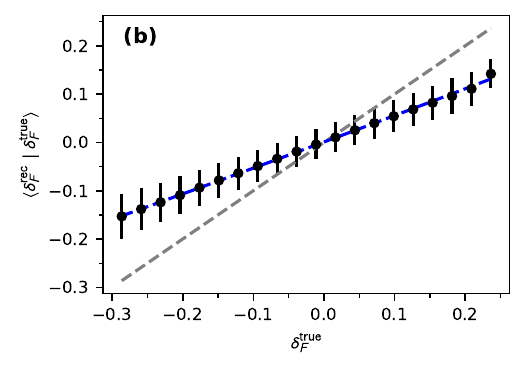}\\\includegraphics[width=3.5in]{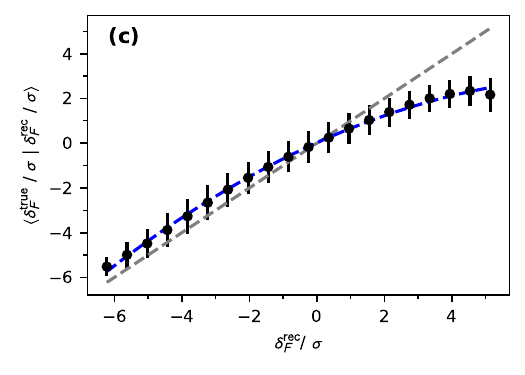}
    \includegraphics[width=3.5in]{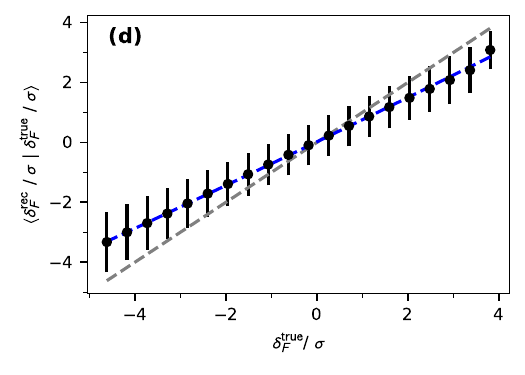}
    \caption{Conditional expectation relations based on the mock surveys, relating the mock-observed $\delta_F^{\rm rec}$ in Wiener-filtered maps to the true value $\delta_F^{\rm true}$. \emph{Panel (a):} Points with error bars show the mean and standard deviation of $\delta_F^{\rm true}$ for voxels in narrow bins of $\delta_F^{\rm rec}$. The blue dashed line is a quadratic fit (see Table~\ref{tab:expectation}). The dashed gray line is the one-to-one line. \emph{Panel (b):} Similarly for $\delta_F^{\rm rec}$ conditional on $\delta_F^{\rm true}$. \emph{Panels (c) and (d):} Like panels (a) and (b), respectively, but with the transmission fluctuations normalized by $\sigma_{\rm map}$.}
    \label{fig:expectations}
\end{figure*}

Our work on the LATIS maps has relied on direct comparisons to the mock-observed maps in order to locate protoclusters, estimate their masses and galaxy overdensities, etc. In this approach, the detailed relationship between the LATIS maps and the true transmission fluctuations is not relevant; all that matters is that the mock observations are faithful, which we have verified by the close agreement between the statistics of the LATIS maps and the mock-observed maps \citep{Newman20,Newman22,Newman25b}. 

\begin{table}
    \centering
    \begin{tabular}{c|ccc}
    Expectation & $c_0$ & $c_1$ & $c_2$ \\\hline
    $\langle \delta_F^{\rm true} | \delta_F^{\rm rec} \rangle$ & -1.147 & 0.919 & 0.001 \\
    $\langle \delta_F^{\rm rec} | \delta_F^{\rm true} \rangle$ & 0.034 & 0.544 & 0.001 \\
    $\langle \delta_F^{\rm true} / \sigma | \delta_F^{\rm rec} / \sigma \rangle$ & -0.039 & 0.682 & 0.023 \\
    $\langle \delta_F^{\rm rec} / \sigma | \delta_F^{\rm true} / \sigma \rangle$ & 0.003 & 0.733 & 0.018 
    \end{tabular}
    \caption{Quadratic fits to the conditional expectations between the true and mock-observed $\delta_F$. Here $\langle y | x \rangle = c_0 x^2 + c_1 x + c_2$, and $\sigma$ denotes $\sigma_{\rm map}$ for the true or mock-observed map according to the quantity being normalized.}
    \label{tab:expectation}
\end{table}

For applications where the quantitative Ly$\alpha$ transmission fluctuation $\delta_F$ is important, it is critical to recognize that there is complex relationship between $\delta_F^{\rm rec}$ and $\delta_F^{\rm true}$. This is examined in Figure~\ref{fig:expectations} using the framework of conditional expectation. Panel (a) shows the conditional expectation $\langle \delta_F^{\rm true} | \delta_F^{\rm rec} \rangle$, estimated as the average value of $\delta_F^{\rm true}$ in the mock survey voxels within a narrow range of $\delta_F^{\rm rec}$; error bars show the standard deviation of $\delta_F^{\rm true}$. The relationship is close to the 1:1 line, showing that $\delta_F^{\rm rec}$ is a nearly unbiased estimator of $\delta_F^{\rm true}$, a key property of the Wiener filter. There are some deviations that can be corrected using the fitting formula in Table~\ref{tab:expectation}. At high $\delta_F^{\rm rec}$, the difference likely arises because $\delta_F^{\rm true}$ has a physical upper limit (corresponding to $F < 1$) that is not enforced by the Wiener filter.

The reverse is not true: panel (b) shows that $\langle \delta_F^{\rm rec} | \delta_F^{\rm true} \rangle$ is far from $\delta_F^{\rm true}$. This can be understood by thinking of the Wiener filter in a Bayesian framework \citep{Pichon01}, in which the signal covariance is a prior. In locations where $|\delta_F^{\rm true}| \gg 0$, the weight of the data may nonetheless be insufficient to pull it away from the zero-centered prior, resulting in a lower amplitude estimate. This is very important to take into account using the fitting formulae in Table~\ref{tab:expectation} if one is comparing an idealized $\delta_F^{\rm true}$, such as a value taken directly from a simulation (rather than a mock observation), to the LATIS maps. 

Another difference between $\delta_F^{\rm rec}$ and $\delta_F^{\rm true}$ is in their amplitudes. The standard deviation $\sigma_{\rm map}$ of the true maps is, on average, 35\% higher than the mock-observed maps. This does not reflect a fundamental discrepancy: the mock-observed and real LATIS maps have compatible $\sigma_{\rm map}$. The amplitude of fluctuations in the Wiener-filtered maps is sensitive to the assumed signal covariance, which was treated approximately \citep{Newman20}. If we normalize $\delta_F^{\rm rec}$ and $\delta_F^{\rm true}$ by their $\sigma_{\rm map}$ values, as we did in nearly all of our LATIS analyses, then $\langle \delta_F^{\rm true} / \sigma~|~\delta_F^{\rm rec} / \sigma \rangle$ maintains a similar behavior, as shown in Figure~\ref{fig:expectations}(c), while $\langle \delta_F^{\rm rec} / \sigma~|~\delta_F^{\rm true} / \sigma \rangle$ becomes closer to one-to-one, as shown in Figure~\ref{fig:expectations}(d).

\section{Discussion and Summary}

\begin{figure}
    \centering
    \includegraphics[width=3.5in]{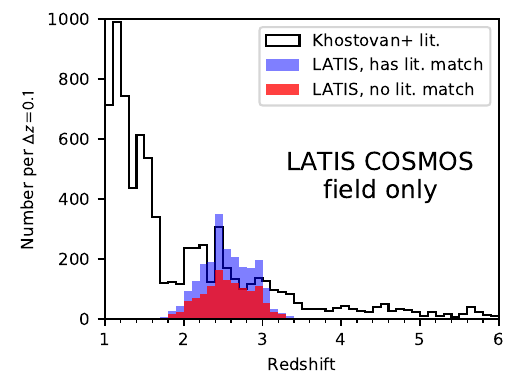}
    \caption{The distribution of confident literature redshifts compiled by \citet{Khostovan25} within the 0.8~deg${}^2$ area of the COSMOS field covered by LATIS, compared to the confident LATIS redshifts (red+blue histograms) and the subset not present in the Khostovan compilation (red).}
    \label{fig:zdist_vs_lit}
\end{figure}

The core LATIS data product is a collection of 7408 Magellan/IMACS spectra of candidate LBGs and QSOs covering the wavelength range 3890--5830~\AA~at $R \sim 1000$ with a typical 12~hr integration time. This data release provides (1) one-dimensional spectra of every target, (2) a catalog of calibrated spectroscopic redshifts, (3) maps of the relative TSR and SSR, (4) Ly$\alpha$ transmission fluctuations in $4.7 \times 10^5$ pixels, and (5) IGM tomography maps covering $4 \times 10^6$~$h^{-3}$~cMpc${}^3$ at $z = 2.2$--2.8 and associated mock surveys.

LATIS constitutes a significant addition to the public database of spectroscopic redshifts at $z \approx 2$--3. In the COSMOS field, \citet{Khostovan25} have assembled 108 sources of spectroscopic redshifts. The surveys providing the largest number of confident (see flags in Section~\ref{sec:z}) redshifts $z = 2$--3 are DESI-ERS \citep{DESI24} with 1139 redshifts, zCOSMOS bright and dark \citep{Lilly07,Lilly09} with 755 redshifts, CLAMATO DR2 \citep{Horowitz22} with 268 redshifts, and MOSDEF \citep{Kriek15} with 210 redshifts. LATIS is larger than all of these, with 1918 confident $z=2$--3 redshifts in COSMOS. Figure~\ref{fig:zdist_vs_lit} shows that this exceeds the 1798 such redshifts amassed in the total prior literature for sources within the 0.8~deg$^2$ LATIS COSMOS field. The other LATIS survey fields, CFHTLS D1 and D4, have been less extensively surveyed, and therefore LATIS marks an even larger increase in public spectroscopic redshifts. 

The LATIS IGM tomography maps are roughly an order of magnitude larger than the first such maps with comparable resolution, which were produced by the CLAMATO survey \citep{Lee14b,Lee18,Horowitz22}, and they are roughly an order of magnitude smaller than those planned for the Subaru Prime Focus Spectrograph (PFS) Galaxy Evolution Survey \citep{Greene22}. The CLAMATO maps cover 0.2~deg${}^2$ of the COSMOS field from $z=2.05$--2.55 \citep{Horowitz22}, covering a volume $4 \times 10^5$~$h^{-3}$~cMpc${}^3$ that is $10\times$ smaller than LATIS. However, the PFS survey plans to cover 12.3 deg${}^2$, about $7\times$ larger than LATIS. All of these surveys have a similar effective sight-line density; volume is the main difference. Reaching smaller scales with higher sight-line densities will likely require future surveys on extremely large telescopes.

\subsection{Data Access}

The LATIS data release is available at \url{http://users.obs.carnegiescience.edu/anewman/latis} and via Zenodo\footnote{\dataset[doi: 10.5281/zenodo.15557327]{\doi{10.5281/zenodo.15557327}}}.

\begin{acknowledgments}
We thank N.~Reddy for sharing his composite LBG spectrum. This paper includes data gathered with the 6.5~m Magellan Telescopes located at Las Campanas Observatory, Chile. We gratefully acknowledge the support of the Observatory staff. A.B.N.~and S.B.~acknowledge support from the National Science Foundation under grant Nos.~2108014 and 2107821, respectively. S.B.~acknowledges funding from NASA ATP 80NSSC22K1897. B.C.L.~acknowledges support from NSF Grant No.~1908422. B.C.L.~is supported by the international Gemini Observatory, a program of NSF NOIRLab, which is managed by the Association of Universities for Research in Astronomy (AURA) under a cooperative agreement with the U.S. National Science Foundation, on behalf of the Gemini partnership of Argentina, Brazil, Canada, Chile, the Republic of Korea, and the United States of America. Based on observations obtained with MegaPrime/MegaCam, a joint project of CFHT and CEA/IRFU, at the Canada–France–Hawaii Telescope (CFHT) which is operated by the National Research Council (NRC) of Canada, the Institut National des Science de l’Univers of the Centre National de la Recherche Scientifique (CNRS) of France, and the University of Hawaii. This work is based in part on data products produced at Terapix, available at the Canadian Astronomy Data Centre as part of the Canada–-France–-Hawaii Telescope Legacy Survey, a collaborative project of NRC and CNRS. 
\end{acknowledgments}

\begin{software}
{\tt scipy} \citep{scipy}, {\tt astropy}  \citep{Astropy13,Astropy18,Astropy22}, {\tt dachshund} (\url{https://github.com/caseywstark/dachshund})
\end{software}

\bibliography{latisdr1}{}
\bibliographystyle{aasjournalv7}

\end{document}